\pdfoutput=1
\documentclass[12pt]{article}%
\usepackage{amsfonts}
\usepackage{mitpress}
\usepackage{amsmath}
\usepackage{amssymb}
\usepackage{graphicx}%
\providecommand{\U}[1]{\protect\rule{.1in}{.1in}}
\newtheorem{theorem}{Theorem}

\newtheorem{corollary}[theorem]{Corollary}

\newtheorem{definition}[theorem]{Definition}
\newtheorem{example}[theorem]{Example}

\newtheorem{notation}[theorem]{Notation}

\newtheorem{remark}[theorem]{Remark}

\newenvironment{proof}[1][Proof]{\noindent\textbf{#1.} }{\ \rule{0.5em}{0.5em}}
\newdimen\dummy
\dummy=\oddsidemargin
\begin{document}

\title{\textbf{UNIVERSAL REGULAR AUTONOMOUS ASYNCHRONOUS SYSTEMS: }$\omega
-$\textbf{LIMIT SETS, INVARIANCE AND BASINS OF ATTRACTION}}
\author{Serban E. Vlad\\Str. Zimbrului, Nr. 3, Bl. PB68, Ap. 11, 410430, Oradea, Romania, mail:
serban\_e\_vlad@yahoo.com, web page: www.serbanvlad.ro}
\maketitle

\begin{abstract}
The asynchronous systems are the non-deterministic real time-binary models of
the asynchronous circuits from electrical engineering. Autonomy means that the
circuits and their models have no input. Regularity means analogies with the
dynamical systems, thus such systems may be considered to be the real time
dynamical systems with a 'vector field' $\Phi:\{0,1\}^{n}\rightarrow
\{0,1\}^{n}.$ Universality refers to the case when the state space of the
system is the greatest possible in the sense of the inclusion. The purpose of
the paper is that of defining, by analogy with the dynamical systems theory,
the $\omega-$limit sets, the invariance and the basins of attraction of the
universal regular autonomous asynchronous systems.

\end{abstract}

\textbf{MSC}: 94C10

\textbf{keywords}: asynchronous system, $\omega-$limit set, invariance, basin
of attraction

\section{Foreword}

Adelina Georgescu founded ROMAI, the Romanian Society of Applied and
Industrial Mathematics in 1992 and I met her in 1993 in Oradea, at the first
Conference on Applied and Industrial Mathematics CAIM. She was a severe woman,
strongly dominant. We got closer in February 2007, when she asked me which are
my mathematical interests -the asynchronous systems- and I asked her which are
her mathematical interests -the dynamical systems. I realized instantly that
the two theories may interact, the first one made on $\mathbf{R}%
\rightarrow\{0,1\}$ functions, the second one made on $\mathbf{R}%
\rightarrow\mathbf{R}$ functions, since the reasoning in asynchronous systems
is often made by following analogies with the real numbers concepts and the
dynamical systems looked to be a real source of inspiration in the sense of
creating analogies. I told her my thoughts and in the months that followed I
received from her by mail many books in dynamical systems written by herself
and by her PhD students. In August that year, near Athens, I presented at a
WSEAS plenary lecture my first work in asynchronous systems considered as
Boolean dynamical systems.

I was deeply impressed by the evolution of her disease and by her
disappearance. She has encouraged me to continue studying the asynchronous
systems, a direction of research that is not popular, she published my papers,
she gave me suggestions of research and consistent bibliography. She will
always remain a model for us, those that had the chance to meet her, with her
idealism, with her wish to construct and with her disappointments, with her
strength and with her fragility. It is a tender pleasure for me to dedicate to
her memory this paper, that is a direct consequence of our friendship.

\section{Introduction}

The mathematical theory of modeling the asynchronous circuits from electrical
engineering has developed in the 50's and the 60's under the name of switching
theory. Afterwards the mathematicians seem to have lost their interest in this
field (or maybe their studies continue, under an unpublished form) and their
work was continued by the engineers. This situation has generated great
theoretical needs in time. In 2007 when our book Asynchronous Systems Theory
was
published\footnote{http://www.romai.ro/lucrari/teoria\_sistemelor\_asincrone\_en.pdf}
the bibliography was poor, consisting mainly in engineering works that give
intuition and we were not sure which MSC it should have, due to its distance
from the general accepted directions of mathematical research. The
'asynchronous systems theory' is a well known syntagm to the engineers,
however the mathematicians do not know it. Professor Valeriu Prepelita helped
in this sense by suggesting, as reviewer, *93-02 Research monographs (systems
and control), 93A05 Axiomatic system theory and 93A10 General systems. We have
adopted in this paper 94C10 Switching theory, application of Boolean algebra,
with the remark that this is a modern point of view of what has been done 50
years ago in switching theory.

Since 2007 our efforts were those of taking the old-newborn asynchronous
systems theory away from its isolation and making it 'socialize' with the
existing systems theory.

The $\mathbf{R}\rightarrow\{0,1\}$ functions give the
deterministic\footnote{'Deterministic' means that each signal is modeled by
exactly one $\mathbf{R}\rightarrow\{0,1\}$ function.} real time-binary models
of the digital electrical signals and they are not studied in literature. An
asynchronous circuit without input, considered as a collection of $n$ signals,
should be deterministically modelled by a function $x:\mathbf{R}%
\rightarrow\{0,1\}^{n}$ called state. We have however several parameters
related with the asynchronous circuit that are either unknown, or perhaps
variable or simply ignored in modeling such as the temperature, the tension of
the mains and the delays the occur in the computation of the Boolean
functions. For this reason, instead of a function $x$ we have in general a set
$X$ of functions $x,$ called state space, or
non-deterministic\footnote{'Non-deterministic' means that each signal is
modeled by several $x_{i}:\mathbf{R}\rightarrow\{0,1\}$ functions or,
equivalently, that each circuit is modeled by several functions $x\in X.$}
autonomous asynchronous system, where each function $x$ represents a
possibility of modeling the circuit. When $X$ is constructed by making use of
a 'vector field' $\Phi:\{0,1\}^{n}\rightarrow\{0,1\}^{n},$ the system $X$ is
called regular. The universal regular autonomous asynchronous systems are the
Boolean dynamical systems and they can be identified with $\Phi$.

\begin{figure}
[ptb]
\begin{center}
\includegraphics[
natheight=0.875200in,
natwidth=1.323200in,
height=0.9072in,
width=1.3578in
]%
{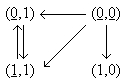}%
\caption{Example of state portrait.}%
\label{attractive3}%
\end{center}
\end{figure}

The dynamic of these systems is described by the so called state portraits.
For example in Figure \ref{attractive3} we have the function $\Phi
:\{0,1\}^{2}\rightarrow\{0,1\}^{2}$
that is defined by the formula $\forall\mu\in\mathbf{B}^{2},(\Phi_{1}(\mu
_{1},\mu_{2}),\Phi_{2}(\mu_{1},\mu_{2}))=(\overline{\mu_{1}}\cup\mu
_{1}\overline{\mu_{2}},\overline{\mu_{1}}\cup\mu_{1}\mu_{2})$ and the arrows
show the increase of time. For any $i\in\{1,2\},$ the coordinate $\mu_{i}$ is
underlined if $\Phi_{i}(\mu_{1},\mu_{2})\neq\mu_{i}$ and it is called
unstable, or enabled, or excited in this case. The coordinates $\mu_{i}$ that
are not underlined satisfy by definition $\Phi_{i}(\mu_{1},\mu_{2})=\mu_{i}$
and are called stable, or disabled, or not excited. Three arrows start from
the point $(0,0)$ where both coordinates are unstable, showing the fact that
$\Phi_{1}(0,0)$ may be computed first, $\Phi_{2}(0,0)$ may be computed first
or $\Phi_{1}(0,0),\Phi_{2}(0,0)$ may be computed simultaneously. Note that the
two possibilities of defining the system, state portrait and formula, are
equivalent. Note also that the system was identified with the function $\Phi$.

The existence of several possibilities of evolution of the system (three
possibilities in $(0,0)$) is the key characteristic of asynchronicity, as
opposed to synchronicity where the coordinates $\Phi_{i}(\mu)$ are always
computed simultaneously, $i\in\{1,...,n\}$ for all $\mu\in\{0,1\}^{n}$ and the
system's run is: $\mu,\Phi(\mu),(\Phi\circ\Phi)(\mu),...,(\Phi\circ
...\circ\Phi)(\mu),...$

Our prezent aim is to show how the well known concepts of $\omega-$limit set,
invariance and basin of attraction from the dynamical systems theory, by real
to binary translation, create asynchronous meanings.

\section{Preliminaries}

\begin{notation}
The set $\mathbf{B}=\{0,1\}$ is the binary Boole algebra, endowed with the
usual algebraical laws and with the discrete topology.
\end{notation}

\begin{definition}
i) The sequence $\alpha:\mathbf{N}\rightarrow\mathbf{B}^{n},\forall
k\in\mathbf{N},\alpha(k)\overset{not}{=}\alpha^{k}$ is called
\textbf{progressive} if the sets $\{k|k\in\mathbf{N},\alpha_{i}^{k}=1\}$ are
infinite for all $i\in\{1,...,n\}.$ We denote the set of the progressive
sequences by $\Pi_{n}.$

ii) $\chi_{A}:\mathbf{R}\rightarrow\mathbf{B}$ is the notation of the
characteristic function of the set $A\subset\mathbf{R}$: $\forall
t\in\mathbf{R},$ $\chi_{A}(t)=\left\{
\begin{array}
[c]{c}%
1,t\in A\\
0,t\notin A
\end{array}
\right.  $ and we also denote by $Seq$ the set of the sequences $t_{0}%
<t_{1}<...<t_{k}<...$ of real numbers that are unbounded from above. The
functions $\rho:\mathbf{R}\rightarrow\mathbf{B}^{n}$ of the form $\forall
t\in\mathbf{R},$%
\begin{equation}
\rho(t)=\alpha^{0}\chi_{\{t_{0}\}}(t)\oplus\alpha^{1}\chi_{\{t_{1}\}}%
(t)\oplus...\oplus\alpha^{k}\chi_{\{t_{k}\}}(t)\oplus... \label{1}%
\end{equation}
where $\alpha\in\Pi_{n}$ and $(t_{k})\in Seq$ are called \textbf{progressive}
and their set is denoted by $P_{n}.$
\end{definition}

\begin{definition}
Let be the function $\Phi:\mathbf{B}^{n}\rightarrow\mathbf{B}^{n}.$ i) For
$\nu\in\mathbf{B}^{n}$ we define $\Phi^{\nu}:\mathbf{B}^{n}\rightarrow
\mathbf{B}^{n}$ by $\forall\mu\in\mathbf{B}^{n},$ $\Phi^{\nu}(\mu
)=(\overline{\nu_{1}}\mu_{1}\oplus\nu_{1}\Phi_{1}(\mu),...,\overline{\nu_{n}%
}\mu_{n}\oplus\nu_{n}\Phi_{n}(\mu)).$ We have denoted with $^{\prime}%
\oplus^{\prime}$ the modulo 2 sum.

ii) The functions $\Phi^{\alpha^{0}...\alpha^{k}}:\mathbf{B}^{n}%
\rightarrow\mathbf{B}^{n}$ are defined for $k\in\mathbf{N}$ and $\alpha
^{0},...,\alpha^{k}\in\mathbf{B}^{n}$ iteratively: $\forall\mu\in
\mathbf{B}^{n},$ $\Phi^{\alpha^{0}...\alpha^{k}\alpha^{k+1}}(\mu)=\Phi
^{\alpha^{k+1}}(\Phi^{\alpha^{0}...\alpha^{k}}(\mu)).$

iii) The function $\Phi^{\rho}:\mathbf{B}^{n}\times\mathbf{R}\rightarrow
\mathbf{B}^{n}$ that is defined in the following way $\Phi^{\rho}(\mu
,t)=\mu\chi_{(-\infty,t_{0})}(t)\oplus\Phi^{\alpha^{0}}(\mu)\chi_{\lbrack
t_{0},t_{1})}(t)\oplus\Phi^{\alpha^{0}\alpha^{1}}(\mu)\chi_{\lbrack
t_{1},t_{2})}(t)\oplus...\oplus\Phi^{\alpha^{0}...\alpha^{k}}(\mu
)\chi_{\lbrack t_{k},t_{k+1})}(t)\oplus...$is called \textbf{flow},
\textbf{motion} or \textbf{orbit} (of $\mu\in\mathbf{B}^{n}$). We have assumed
that $\rho\in P_{n}$ is like at (\ref{1}).

iv) The set $Or_{\rho}(\mu)=\{\Phi^{\rho}(\mu,t)|t\in\mathbf{R}\}$is also
called \textbf{orbit} (of $\mu$).
\end{definition}

\begin{remark}
The function $\Phi^{\nu}$ shows how an asynchronous iteration of $\Phi$ is
made: for any $i\in\{1,...,n\},$ if $\nu_{i}=0$ then $\Phi_{i}$ is not
computed, since $\Phi_{i}^{\nu}(\mu)=\mu_{i}$ and if $\nu_{i}=1$ then
$\Phi_{i}$ is computed, since $\Phi_{i}^{\nu}(\mu)=\Phi_{i}(\mu).$

The definition of $\Phi^{\alpha^{0}...\alpha^{k}}$ generalizes this idea to an
arbitrary number $k+1$ of asynchronous iterations, with the supplementary
request that each coordinate $\Phi_{i}$ is computed infinitely many times in
the sequence $\mu,\Phi^{\alpha^{0}}(\mu),\Phi^{\alpha^{0}\alpha^{1}}%
(\mu),...,\Phi^{\alpha^{0}...\alpha^{k}}(\mu),...$ whenever $\alpha\in\Pi
_{n}.$

The sequences $(t_{k})\in Seq$ make the pass from the discrete time
$\mathbf{N}$ to the continuous time $\mathbf{R}$ and each $\rho\in P_{n}$
shows, in addition to $\alpha\in\Pi_{n}$, the time instants $t_{k}$ when
$\Phi$ is computed (asynchronously). Thus $\Phi^{\rho}(\mu,t),t\in\mathbf{R}$
is the continuous time computation of the sequence $\mu,$ $\Phi^{\alpha^{0}%
}(\mu),$ $\Phi^{\alpha^{0}\alpha^{1}}(\mu),$ $...,$ $\Phi^{\alpha^{0}%
...\alpha^{k}}(\mu),$ $...$ made in the following way: if $t<t_{0}$ nothing is
computed, if $t\in\lbrack t_{0},t_{1}),$ $\Phi^{\alpha^{0}}(\mu)$ is computed,
if $t\in\lbrack t_{1},t_{2}),\Phi^{\alpha^{0}\alpha^{1}}(\mu)$ is computed,
..., if $t\in\lbrack t_{k},t_{k+1}),\Phi^{\alpha^{0}...\alpha^{k}}(\mu)$ is
computed, ...

When $\alpha$ runs in $\Pi_{n}$ and $(t_{k})$ runs in $Seq$ we get the
'unbounded delay model' of computation of the Boolean function $\Phi$,
represented in discrete time by the sequences $\mu,\Phi^{\alpha^{0}}(\mu
),\Phi^{\alpha^{0}\alpha^{1}}(\mu),...,\Phi^{\alpha^{0}...\alpha^{k}}%
(\mu),...$ and in continuous time by the orbits $\Phi^{\rho}(\mu,t)$
respectively. We shall not insist on the non-formalized way that the engineers
describe this model; we just mention that the 'unbounded delay model' is a
reasonable way of starting the analysis of a circuit in which the delays
occurring in the computation of the Boolean functions $\Phi$ are arbitrary
positive numbers. If we restrict suitably the ranges of $\alpha$ and $(t_{k})$
we get the 'bounded delay model' of computation of $\Phi$ and if both $\alpha
$, $(t_{k})$ are fixed, then we obtain the 'fixed delay model' of computation
of $\Phi,$ determinism.
\end{remark}

\begin{theorem}
\label{The10}Let $\alpha\in\Pi_{n},(t_{k})\in Seq$ be arbitrary and the
function $\rho(t)=\alpha^{0}\chi_{\{t_{0}\}}(t)\oplus\alpha^{1}\chi
_{\{t_{1}\}}(t)\oplus...\oplus\alpha^{k}\chi_{\{t_{k}\}}(t)\oplus...,\rho\in
P_{n}.$ The following statements are true:

a) $\{\alpha^{k}|k\geq k_{1}\}\in\Pi_{n}$ for any $k_{1}\in\mathbf{N};$

b) $(t_{k})\cap(t^{\prime},\infty)\in Seq$ for any $t^{\prime}\in\mathbf{R};$

c) $\rho\chi_{(t^{\prime},\infty)}\in P_{n}$ for any $t^{\prime}\in
\mathbf{R};$

d) $\forall\mu\in\mathbf{B}^{n},\forall\mu^{\prime}\in\mathbf{B}^{n},\forall
t^{\prime}\in\mathbf{R},\Phi^{\rho}(\mu,t^{\prime})=\mu^{\prime}%
\Longrightarrow\forall t\geq t^{\prime},\Phi^{\rho}(\mu,t)=\Phi^{\rho
\chi_{(t^{\prime},\infty)}}(\mu^{\prime},t).$
\end{theorem}

\begin{proof}
a) If $\{k|k\in\mathbf{N},\alpha_{i}^{k}=1\}$ is infinite, then $\{k|k\geq
k_{1},\alpha_{i}^{k}=1\}$ is also infinite, $\forall i\in\{1,...,n\}.$

b) If $t_{0}<t_{1}<t_{2}<...$ is unbounded from above, then any sequence of
the form $t_{k_{1}}<t_{k_{1}+1}<t_{k_{1}+2}<...$ is unbounded from above,
$k_{1}\in\mathbf{N}$.

c) This is a consequence of a) and b).

d) We presume that $t^{\prime}<t_{0}.$ In this situation $\mu=\mu^{\prime
},\rho=\rho\chi_{(t^{\prime},\infty)}$ and the statement is obvious, so that
we may assume now that $t^{\prime}\geq t_{0}.$ In this case, some $k_{1}%
\in\mathbf{N}$ exists with $t^{\prime}\in\lbrack t_{k_{1}},t_{k_{1}+1})$ and
$\mu^{\prime}=\Phi^{\alpha^{0}...\alpha^{k_{1}}}(\mu).$ Because%
\[
\rho\chi_{(t^{\prime},\infty)}(t)=\alpha^{k_{1}+1}\chi_{\{t_{k_{1}+1}%
\}}(t)\oplus\alpha^{k_{1}+2}\chi_{\{t_{k_{1}+2}\}}(t)\oplus...,
\]%
\[
\Phi^{\rho\chi_{(t^{\prime},\infty)}}(\mu^{\prime},t)=\mu^{\prime}%
\chi_{(-\infty,t_{k_{1}+1})}(t)\oplus\Phi^{\alpha^{k_{1}+1}}(\mu^{\prime}%
)\chi_{\lbrack t_{k_{1}+1},t_{k_{1}+2})}(t)
\]%
\[
\oplus\Phi^{\alpha^{k_{1}+1}\alpha^{k_{1}+2}}(\mu^{\prime})\chi_{\lbrack
t_{k_{1}+2},t_{k_{1}+3})}(t)\oplus...
\]
we get

$\forall t\in\lbrack t^{\prime},t_{k_{1}+1}),$%
\[
\Phi^{\rho}(\mu,t)=\Phi^{\alpha^{0}...\alpha^{k_{1}}}(\mu),
\]%
\[
\Phi^{\rho\chi_{(t^{\prime},\infty)}}(\mu^{\prime},t)=\mu^{\prime}%
=\Phi^{\alpha^{0}...\alpha^{k_{1}}}(\mu);
\]

$\forall t\in\lbrack t_{k_{1}+1},t_{k_{1}+2}),$%
\[
\Phi^{\rho}(\mu,t)=\Phi^{\alpha^{0}...\alpha^{k_{1}}\alpha^{k_{1}+1}}(\mu),
\]%
\[
\Phi^{\rho\chi_{(t^{\prime},\infty)}}(\mu^{\prime},t)=\Phi^{\alpha^{k_{1}+1}%
}(\mu^{\prime})=\Phi^{\alpha^{k_{1}+1}}(\Phi^{\alpha^{0}...\alpha^{k_{1}}}%
(\mu))=\Phi^{\alpha^{0}...\alpha^{k_{1}}\alpha^{k_{1}+1}}(\mu);
\]%
\[
...
\]
The statement of the Theorem holds.
\end{proof}

\begin{theorem}
\label{The11}Let be $\mu\in\mathbf{B}^{n},\rho\in P_{n}$ and $\tau
\in\mathbf{R}.$ The function $\rho^{\prime}(t)=\rho(t-\tau)$ is progressive
and we have $\Phi^{\rho^{\prime}}(\mu,t)=\Phi^{\rho}(\mu,t-\tau).$
\end{theorem}

\begin{proof}
We put $\rho$ under the form%
\[
\rho(t)=\alpha^{0}\chi_{\{t_{0}\}}(t)\oplus...\oplus\alpha^{k}\chi_{\{t_{k}%
\}}(t)\oplus...,
\]
$\alpha\in\Pi_{n},(t_{k})\in Seq$ and we note that%
\[
\rho^{\prime}(t)=\rho(t-\tau)=\alpha^{0}\chi_{\{t_{0}+\tau\}}(t)\oplus
...\oplus\alpha^{k}\chi_{\{t_{k}+\tau\}}(t)\oplus...
\]
where $(t_{k}+\tau)\in Seq.$ We infer%
\[
\Phi^{\rho^{\prime}}(\mu,t)=\mu\chi_{(-\infty,t_{0}+\tau)}(t)\oplus
\Phi^{\alpha^{0}}(\mu)\chi_{\lbrack t_{0}+\tau,t_{1}+\tau)}(t)\oplus...
\]%
\[
...\oplus\Phi^{\alpha^{0}...\alpha^{k}}(\mu)\chi_{\lbrack t_{k}+\tau
,t_{k+1}+\tau)}(t)\oplus...=\Phi^{\rho}(\mu,t-\tau).
\]

\end{proof}

\begin{definition}
The \textbf{universal regular autonomous asynchronous system} that is
generated by $\Phi:\mathbf{B}^{n}\rightarrow\mathbf{B}^{n}$ is by definition
$\Xi_{\Phi}=\{\Phi^{\rho}(\mu,\cdot)|\mu\in\mathbf{B}^{n},\rho\in P_{n}\};$
any $x(t)=\Phi^{\rho}(\mu,t)$ is called \textbf{state} (of $\Xi_{\Phi}$),
$\mu$ is called \textbf{initial value} (of $x$), or \textbf{initial state} (of
$\Xi_{\Phi}$) and $\Phi$ is called \textbf{generator function} (of $\Xi_{\Phi
}$).
\end{definition}

\begin{remark}
The asynchronous systems are non-deterministic in general, due to the
uncertainties that occur in the modeling of the asynchronous circuits.
Non-determinism is produced, in the case of $\Xi_{\Phi}$, by the fact that the
initial state $\mu$ and the way $\rho$ of iterating $\Phi$ are not known.
\end{remark}

\begin{definition}
Let $v:\mathbf{N}\rightarrow\mathbf{B}^{n},x:\mathbf{R}\rightarrow
\mathbf{B}^{n}$ be some functions. If $\exists k^{\prime}\in\mathbf{N},\forall
k\geq k^{\prime},v(k)=v(k^{\prime}),$ we say that \textbf{the limit}
$\underset{k\rightarrow\infty}{\lim}v(k)$ \textbf{exists} and we use the
notation $\underset{k\rightarrow\infty}{\lim}v(k)=v(k^{\prime}).$ Similarly,
if $\exists t^{\prime}\in\mathbf{R},\forall t\geq t^{\prime},x(t)=x(t^{\prime
}),$we say that \textbf{the limit} $\underset{t\rightarrow\infty}{\lim}x(t)$
\textbf{exists} and we denote $\underset{t\rightarrow\infty}{\lim
}x(t)=x(t^{\prime}).$ Sometimes $\underset{k\rightarrow\infty}{\lim
}v(k),\underset{t\rightarrow\infty}{\lim}x(t)$ are called the \textbf{final
values} of $v,x.$
\end{definition}

\begin{theorem}
\label{The15}\cite{bib7} $\forall\mu\in\mathbf{B}^{n},\forall\mu^{\prime}%
\in\mathbf{B}^{n},\forall\rho\in P_{n},\underset{t\rightarrow\infty}{\lim}%
\Phi^{\rho}(\mu,t)=\mu^{\prime}\Longrightarrow\Phi(\mu^{\prime})=\mu^{\prime
},$ if the final value of $\Phi^{\rho}(\mu,\cdot)$ exists, it is a fixed point
of $\Phi$.
\end{theorem}

\begin{proof}
Let $\mu\in\mathbf{B}^{n},\mu^{\prime}\in\mathbf{B}^{n},\rho\in P_{n}$ be
arbitrary and fixed. The hypothesis states the existence of $t^{\prime}%
\in\mathbf{R}$ with%
\[
\forall t\geq t^{\prime},\Phi^{\rho}(\mu,t)=\mu^{\prime}%
\]
thus, from Theorem \ref{The10} d),%
\[
\forall t\geq t^{\prime},\Phi^{\rho\chi_{(t^{\prime},\infty)}}(\mu^{\prime
},t)=\mu^{\prime}.
\]
We infer that $\forall i\in\{1,...,n\},\exists t^{\prime\prime}>t^{\prime}$
such that%
\[
\rho_{i}(t^{\prime\prime})=\rho_{i}\chi_{(t^{\prime},\infty)}(t^{\prime\prime
})=1,
\]%
\[
\Phi_{i}^{\rho\chi_{(t^{\prime},\infty)}}(\mu^{\prime},t^{\prime\prime}%
)=\Phi_{i}(\mu^{\prime})=\mu_{i}^{\prime}.
\]

\end{proof}

\begin{theorem}
\label{The16}\cite{bib7} $\forall\mu\in\mathbf{B}^{n},\forall\mu^{\prime}%
\in\mathbf{B}^{n},\forall\rho\in P_{n},(\Phi(\mu^{\prime})=\mu^{\prime
}\;and\;\exists t^{\prime}\in\mathbf{R},\Phi^{\rho}(\mu,t^{\prime}%
)=\mu^{\prime})\Longrightarrow\forall t\geq t^{\prime},\Phi^{\rho}(\mu
,t)=\mu^{\prime},$meaning that if the fixed point $\mu^{\prime}$ of $\Phi$ is
accessible, then it is the final value of $\Phi^{\rho}(\mu,\cdot).$
\end{theorem}

\begin{proof}
Let $\mu\in\mathbf{B}^{n},\mu^{\prime}\in\mathbf{B}^{n},\rho\in P_{n}$ be
arbitrary and fixed. From the hypothesis and Theorem \ref{The10} d) we infer%
\[
\forall t\geq t^{\prime},\Phi^{\rho}(\mu,t)=\Phi^{\rho\chi_{(t^{\prime}%
,\infty)}}(\mu^{\prime},t)
\]
thus $\forall i\in\{1,...,n\},\exists\varepsilon>0,\forall t\in\lbrack
t^{\prime},t^{\prime}+\varepsilon),\Phi_{i}^{\rho\chi_{(t^{\prime},\infty)}%
}(\mu^{\prime},t)$ can take one of the values $\mu_{i}^{\prime}$ and $\Phi
_{i}(\mu^{\prime}).$ But $\mu_{i}^{\prime}=\Phi_{i}(\mu^{\prime}),$ wherefrom
the previous property takes place for arbitrary $\varepsilon$ and%
\[
\forall t\geq t^{\prime},\Phi^{\rho}(\mu,t)=\mu^{\prime}.
\]

\end{proof}

\begin{corollary}
\label{Cor17}$\forall\mu\in\mathbf{B}^{n},\forall\rho\in P_{n},\Phi(\mu
)=\mu\Longrightarrow\forall t\in\mathbf{R},\Phi^{\rho}(\mu,t)=\mu.$
\end{corollary}

\begin{proof}
From Theorem \ref{The16}, with $\mu=\mu^{\prime},$ where $t^{\prime}$ may be
chosen such that $\forall t<t^{\prime},\rho(t)=0.$
\end{proof}

\section{$\omega-$limit sets}

\begin{definition}
For $\mu\in\mathbf{B}^{n}$ and $\rho\in P_{n},$ the set $\omega_{\rho}%
(\mu)=\{\mu^{\prime}|\mu^{\prime}\in\mathbf{B}^{n},\exists(t_{k})\in
Seq,\underset{k\rightarrow\infty}{\lim}\Phi^{\rho}(\mu,t_{k})=\mu^{\prime}\}$
is called the $\omega-$\textbf{limit set} of the orbit $\Phi^{\rho}(\mu
,\cdot).$
\end{definition}

\begin{remark}
The previous definition agrees with the usual definitions of the $\omega
-$limit sets of the real time or discrete time dynamical systems see
\cite{bib3} page 5, \cite{bib2} page 26, \cite{bib4} page 20.
\end{remark}

\begin{example}
In Figure \ref{attractor41}, we consider%
\begin{figure}
[ptb]
\begin{center}
\includegraphics[
natheight=0.864800in,
natwidth=1.208100in,
height=0.8968in,
width=1.2427in
]%
{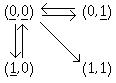}%
\caption{$\exists\rho\in P_{2},\omega_{\rho}((1,0))=\{(0,0),(0,1)\}$ and
$\exists\rho^{\prime}\in P_{2},\omega_{\rho^{\prime}}((1,0))=\{(1,1)\}$}%
\label{attractor41}%
\end{center}
\end{figure}
\[
\rho(t)=(1,1)\chi_{\{0\}}(t)\oplus(0,1)\chi_{\{1\}}(t)\oplus(1,1)\chi
_{\{2\}}(t)\oplus(0,1)\chi_{\{3\}}(t)\oplus...,
\]%
\[
\rho^{\prime}(t)=(1,1)\chi_{\{0\}}(t)\oplus(1,1)\chi_{\{1\}}(t)\oplus
(1,1)\chi_{\{2\}}(t)\oplus...
\]
and we have%
\[
\Phi^{\rho}((1,0),t)=(1,0)\chi_{(-\infty,0)}(t)\oplus(0,0)\chi_{\lbrack
0,1)}(t)\oplus(0,1)\chi_{\lbrack1,2)}(t)
\]%
\[
\oplus(0,0)\chi_{\lbrack2,3)}(t)\oplus(0,1)\chi_{\lbrack3,4)}(t)\oplus...,
\]%
\[
\Phi^{\rho^{\prime}}((1,0),t)=(1,0)\chi_{(-\infty,0)}(t)\oplus(0,0)\chi
_{\lbrack0,1)}(t)\oplus(1,1)\chi_{\lbrack1,\infty)}(t),
\]
thus $\omega_{\rho}((1,0))=\{(0,0),(0,1)\},\omega_{\rho^{\prime}%
}((1,0))=\{(1,1)\}.$
\end{example}

\begin{theorem}
\label{The20_}For any $\mu\in\mathbf{B}^{n}$ and any $\rho\in P_{n},$ we have:

a) $\omega_{\rho}(\mu)\neq\emptyset;$

b) $\forall t^{\prime}\in\mathbf{R},$ $\omega_{\rho}(\mu)\subset\{\Phi^{\rho
}(\mu,t)|t\geq t^{\prime}\}\subset Or_{\rho}(\mu);$

c) $\exists t^{\prime}\in\mathbf{R},\omega_{\rho}(\mu)=\{\Phi^{\rho}%
(\mu,t)|t\geq t^{\prime}\}$ and any $t^{\prime\prime}\geq t^{\prime}$ fulfills
$\omega_{\rho}(\mu)=\{\Phi^{\rho}(\mu,t)|t\geq t^{\prime\prime}\};$

d) $\forall t^{\prime}\in\mathbf{R},\forall t^{\prime\prime}\geq t^{\prime
},\{\Phi^{\rho}(\mu,t)|t\geq t^{\prime}\}=\{\Phi^{\rho}(\mu,t)|t\geq
t^{\prime\prime}\}$ implies $\omega_{\rho}(\mu)=\{\Phi^{\rho}(\mu,t)|t\geq
t^{\prime}\};$

e) we presume that $\omega_{\rho}(\mu)=\{\Phi^{\rho}(\mu,t)|t\geq t^{\prime
}\},t^{\prime}\in\mathbf{R}.$ Then $\forall\mu^{\prime}\in\omega_{\rho}%
(\mu),\forall t^{\prime\prime}\geq t^{\prime},$ if $\Phi^{\rho}(\mu
,t^{\prime\prime})=\mu^{\prime}$ we get $\omega_{\rho}(\mu)=\{\Phi^{\rho
\chi_{(t^{\prime\prime},\infty)}}(\mu^{\prime},t)|t\geq t^{\prime\prime
}\}=Or_{\rho\chi_{(t^{\prime\prime},\infty)}}(\mu^{\prime})=\omega_{\rho
\chi_{(t^{\prime\prime},\infty)}}(\mu^{\prime}).$
\end{theorem}

\begin{proof}
We put $\rho\in P_{n}$ under the form%
\[
\rho(t)=\alpha^{0}\chi_{\{t_{0}\}}(t)\oplus...\oplus\alpha^{k}\chi_{\{t_{k}%
\}}(t)\oplus...
\]
where $\alpha\in\Pi_{n}$ and $(t_{k})\in Seq.$ We ask, without loosing the
generality, that $\alpha^{0}=(0,...,0)\in\mathbf{B}^{n},$ hence $\Phi^{\rho
}(\mu,t_{0})=\mu$ and $Or_{\rho}(\mu)=\{\Phi^{\rho}(\mu,t_{k})|k\in
\mathbf{N}\}.$

a) If $Or_{\rho}(\mu)=\{\mu^{1},...,\mu^{p}\},p\in\{1,...,2^{n}\},$ we denote
with $I_{1},...,I_{p}\subset\mathbf{N}$ the sets%
\[
I_{j}=\{k|k\in\mathbf{N},\Phi^{\rho}(\mu,t_{k})=\mu^{j}\},j=\overline{1,p}.
\]
Because $I_{1}\cup...\cup I_{p}=\mathbf{N}$, some of these sets are infinite,
let them be without loosing the generality $I_{1},...,I_{p^{\prime}}%
,p^{\prime}\leq p.$ We infer $\omega_{\rho}(\mu)=\{\mu^{1},...,\mu^{p^{\prime
}}\}.$

b) For $t^{\prime}\in\mathbf{R},$ we define%
\[
k_{1}=\left\{
\begin{array}
[c]{c}%
0,t^{\prime}<t_{0}\\
k,t^{\prime}\in\lbrack t_{k},t_{k+1})
\end{array}
\right.
\]
and we obtain%
\[
\omega_{\rho}(\mu)=\{\mu^{1},...,\mu^{p^{\prime}}\}=\{\Phi^{\rho}(\mu
,t_{k})|k\in I_{1}\cup...\cup I_{p^{\prime}}\}
\]%
\[
=\{\Phi^{\rho}(\mu,t_{k})|k\in(I_{1}\cup...\cup I_{p^{\prime}})\cap\lbrack
k_{1},\infty)\}
\]%
\[
\subset\{\Phi^{\rho}(\mu,t_{k})|k\in(I_{1}\cup...\cup I_{p})\cap\lbrack
k_{1},\infty)\}=\{\Phi^{\rho}(\mu,t)|t\geq t^{\prime}\}
\]%
\[
\subset\{\Phi^{\rho}(\mu,t_{k})|k\in I_{1}\cup...\cup I_{p}\}=\{\mu
^{1},...,\mu^{p}\}=Or_{\rho}(\mu).
\]

c) If $p^{\prime}=p,$ then $\forall t^{\prime}\in\mathbf{R},$ $\omega_{\rho
}(\mu)=\{\Phi^{\rho}(\mu,t)|t\geq t^{\prime}\}=Or_{\rho}(\mu)$ from b) and the
property holds, thus we can assume that $p^{\prime}<p.$ In this case we
define
\begin{align*}
k^{\prime\prime}  &  =\min\{k|k\in\mathbf{N},\forall k^{\prime}\geq
k,k^{\prime}\in I_{1}\cup...\cup I_{p^{\prime}}\}\\
&  =1+\max(I_{p^{\prime}+1}\cup...\cup I_{p})
\end{align*}
for which we have%
\[
(I_{p^{\prime}+1}\cup...\cup I_{p})\cap\lbrack k^{\prime\prime},\infty
)=\emptyset
\]
and $t^{\prime}=t_{k^{\prime\prime}}$ fulfills
\[
\omega_{\rho}(\mu)=\{\mu^{1},...,\mu^{p^{\prime}}\}=\{\Phi^{\rho}(\mu
,t_{k})|k\in I_{1}\cup...\cup I_{p^{\prime}}\}
\]%
\[
=\{\Phi^{\rho}(\mu,t_{k})|k\in(I_{1}\cup...\cup I_{p^{\prime}})\cap\lbrack
k^{\prime\prime},\infty)\}
\]%
\[
=\{\Phi^{\rho}(\mu,t_{k})|k\in(I_{1}\cup...\cup I_{p})\cap\lbrack
k^{\prime\prime},\infty)\}=\{\Phi^{\rho}(\mu,t)|t\geq t^{\prime}\};
\]
any $t^{\prime\prime}\geq t^{\prime}$ gives%
\[
\omega_{\rho}(\mu)\overset{b)}{\subset}\{\Phi^{\rho}(\mu,t)|t\geq
t^{\prime\prime}\}\subset\{\Phi^{\rho}(\mu,t)|t\geq t^{\prime}\}=\omega_{\rho
}(\mu).
\]

d) Let be $t^{\prime}\in\mathbf{R}$ such that $\forall t^{\prime\prime}\geq
t^{\prime},$%
\begin{equation}
\{\Phi^{\rho}(\mu,t)|t\geq t^{\prime}\}=\{\Phi^{\rho}(\mu,t)|t\geq
t^{\prime\prime}\} \label{omega1}%
\end{equation}
and we claim that in this case we have%
\begin{equation}
\forall\mu^{\prime}\in\{\Phi^{\rho}(\mu,t)|t\geq t^{\prime}\},\exists
(t_{k}^{\prime})\in Seq,\forall k\in\mathbf{N},\Phi^{\rho}(\mu,t_{k}^{\prime
})=\mu^{\prime}. \label{omega2}%
\end{equation}
We assume against all reason that (\ref{omega2}) is false, meaning that%
\[
\exists\mu^{\prime}\in\{\Phi^{\rho}(\mu,t)|t\geq t^{\prime}\},\text{ the set
}\{t_{k}|k\in\mathbf{N},\Phi^{\rho}(\mu,t_{k})=\mu^{\prime}\}\text{ is
finite.}%
\]
Then $\exists t^{\prime\prime}>\max\{\max\{t_{k}|k\in\mathbf{N},\Phi^{\rho
}(\mu,t_{k})=\mu^{\prime}\},t^{\prime}\}$ that fulfills $\mu^{\prime}\in
\{\Phi^{\rho}(\mu,t)|t\geq t^{\prime}\}\setminus\{\Phi^{\rho}(\mu,t)|t\geq
t^{\prime\prime}\},$ contradiction with (\ref{omega1}). The truth of
(\ref{omega2}) shows that $\mu^{\prime}\in\omega_{\rho}(\mu),$ i.e.
$\{\Phi^{\rho}(\mu,t)|t\geq t^{\prime}\}\subset\omega_{\rho}(\mu).$ For all
$t^{\prime\prime}\geq t^{\prime}$ we have then%
\[
\omega_{\rho}(\mu)\overset{b)}{\subset}\{\Phi^{\rho}(\mu,t)|t\geq
t^{\prime\prime}\}=\{\Phi^{\rho}(\mu,t)|t\geq t^{\prime}\}\subset\omega_{\rho
}(\mu).
\]

e) We note that for $t^{\prime\prime}\geq t^{\prime}$ and $\Phi^{\rho}%
(\mu,t^{\prime\prime})=\mu^{\prime}$ we can write%
\[
\omega_{\rho}(\mu)=\{\Phi^{\rho}(\mu,t)|t\geq t^{\prime}\}\overset{c)}%
{=}\{\Phi^{\rho}(\mu,t)|t\geq t^{\prime\prime}\}
\]%
\[
\overset{Theorem\;\ref{The10}\;d)}{=}\{\Phi^{\rho\chi_{(t^{\prime\prime
},\infty)}}(\mu^{\prime},t)|t\geq t^{\prime\prime}\}=\{\Phi^{\rho
\chi_{(t^{\prime\prime},\infty)}}(\mu^{\prime},t)|t\in\mathbf{R}\}
\]%
\[
=Or_{\rho\chi_{(t^{\prime\prime},\infty)}}(\mu^{\prime}).
\]
The fact that $\forall t^{\prime\prime\prime}\geq t^{\prime\prime},$%
\[
\{\Phi^{\rho\chi_{(t^{\prime\prime},\infty)}}(\mu^{\prime},t)|t\geq
t^{\prime\prime}\}\overset{Theorem\;\ref{The10}\;d)}{=}\{\Phi^{\rho}%
(\mu,t)|t\geq t^{\prime\prime}\}\overset{c)}{=}\{\Phi^{\rho}(\mu,t)|t\geq
t^{\prime}\}
\]%
\[
\overset{c)}{=}\{\Phi^{\rho}(\mu,t)|t\geq t^{\prime\prime\prime}%
\}\overset{Theorem\;\ref{The10}\;d)}{=}\{\Phi^{\rho\chi_{(t^{\prime\prime
},\infty)}}(\mu^{\prime},t)|t\geq t^{\prime\prime\prime}\}
\]
shows, by taking into account d), that%
\[
\{\Phi^{\rho\chi_{(t^{\prime\prime},\infty)}}(\mu^{\prime},t)|t\geq
t^{\prime\prime}\}=\omega_{\rho\chi_{(t^{\prime\prime},\infty)}}(\mu^{\prime
}).
\]

\end{proof}

\begin{remark}
\label{Rem23}If in Theorem \ref{The20_} e) we take $t^{\prime\prime}%
\in\mathbf{R}$ arbitrarily, the equation%
\begin{equation}
\omega_{\rho}(\mu)=\omega_{\rho\chi_{(t^{\prime\prime},\infty)}}(\Phi^{\rho
}(\mu,t^{\prime\prime})) \label{asterisc}%
\end{equation}
is still true. Indeed, for sufficiently great $t^{\prime\prime\prime}$, the
terms in (\ref{asterisc}) are equal with%
\[
\{\Phi^{\rho}(\mu,t)|t\geq t^{\prime\prime\prime}\}=\{\Phi^{\rho
\chi_{(t^{\prime\prime},\infty)}}(\Phi^{\rho}(\mu,t^{\prime\prime}),t)|t\geq
t^{\prime\prime\prime}\}.
\]

\end{remark}

\begin{theorem}
\label{The20}For arbitrary $\mu\in\mathbf{B}^{n}$,$\rho\in P_{n}$ the
following statements are true:

a) $\underset{t\rightarrow\infty}{\lim}\Phi^{\rho}(\mu,t)$ exists
$\Longleftrightarrow card(\omega_{\rho}(\mu))=1;$

b) if $\exists\mu^{\prime}\in\mathbf{B}^{n},\omega_{\rho}(\mu)=\{\mu^{\prime
}\},$ then $\underset{t\rightarrow\infty}{\lim}\Phi^{\rho}(\mu,t)=\mu^{\prime
}$ and $\Phi(\mu^{\prime})=\mu^{\prime};$

c) if $\exists\mu^{\prime}\in\mathbf{B}^{n},\Phi(\mu^{\prime})=\mu^{\prime}$
and $\mu^{\prime}\in Or_{\rho}(\mu),$ then $\omega_{\rho}(\mu)=\{\mu^{\prime
}\}.$
\end{theorem}

\begin{proof}
a) Let $\mu\in\mathbf{B}^{n}$,$\rho\in P_{n}$ be arbitrary. We get%
\[
\underset{t\rightarrow\infty}{\lim}\Phi^{\rho}(\mu,t)\text{ exists}%
\Longleftrightarrow\exists\mu^{\prime}\in\mathbf{B}^{n},\exists t^{\prime}%
\in\mathbf{R},\forall t\geq t^{\prime},\Phi^{\rho}(\mu,t)=\mu^{\prime}%
\]%
\[
\Longleftrightarrow\exists\mu^{\prime}\in\mathbf{B}^{n},\exists t^{\prime}%
\in\mathbf{R},\{\Phi^{\rho}(\mu,t)|t\geq t^{\prime}\}=\{\mu^{\prime}\}
\]%
\[
\Longleftrightarrow\exists\mu^{\prime}\in\mathbf{B}^{n},\omega_{\rho}%
(\mu)=\{\mu^{\prime}\}\Longleftrightarrow card(\omega_{\rho}(\mu))=1.
\]

b) We assume that $\exists\mu^{\prime}\in\mathbf{B}^{n},\omega_{\rho}%
(\mu)=\{\mu^{\prime}\},$ i.e. $\exists\mu^{\prime}\in\mathbf{B}^{n},\exists
t^{\prime}\in\mathbf{R},\{\Phi^{\rho}(\mu,t)|t\geq t^{\prime}\}=\{\mu^{\prime
}\}$ in other words $\underset{t\rightarrow\infty}{\lim}\Phi^{\rho}(\mu
,t)=\mu^{\prime}.$ The fact that $\Phi(\mu^{\prime})=\mu^{\prime}$ results
from Theorem \ref{The15}.

c) This is a consequence of Theorem \ref{The16}.
\end{proof}

\begin{theorem}
\label{The22}Let be $\mu\in\mathbf{B}^{n},\rho\in P_{n},\tau\in\mathbf{R}.$
The function $\rho^{\prime}\in P_{n},\rho^{\prime}(t)=\rho(t-\tau)$ fulfills
$\omega_{\rho}(\mu)=\omega_{\rho^{\prime}}(\mu).$
\end{theorem}

\begin{proof}
We use Theorem \ref{The11} and we infer the existence of $t^{\prime}%
\in\mathbf{R}$ such that%
\[
\omega_{\rho}(\mu)=\{\Phi^{\rho}(\mu,t)|t\geq t^{\prime}\}=\{\Phi^{\rho}%
(\mu,t-\tau)|t-\tau\geq t^{\prime}\}
\]%
\[
=\{\Phi^{\rho^{\prime}}(\mu,t)|t\geq t^{\prime}+\tau\}=\omega_{\rho^{\prime}%
}(\mu).
\]

\end{proof}

\section{P-invariant and n-invariant sets}

\begin{theorem}
\label{The14}We consider the function $\Phi:\mathbf{B}^{n}\rightarrow
\mathbf{B}^{n}$ and let be the set $A\in P^{\ast}(\mathbf{B}^{n}).$ For any
$\mu\in A,$ the following properties are equivalent%
\begin{equation}
\exists\rho\in P_{n},Or_{\rho}(\mu)\subset A, \label{inv2}%
\end{equation}%
\begin{equation}
\exists\rho\in P_{n},\forall t\in\mathbf{R},\Phi^{\rho}(\mu,t)\in A,
\label{inv1__}%
\end{equation}%
\begin{equation}
\exists\alpha\in\Pi_{n},\forall k\in\mathbf{N},\Phi^{\alpha^{0}...\alpha^{k}%
}(\mu)\in A \label{inv3__}%
\end{equation}
and the following properties are also equivalent%
\begin{equation}
\forall\rho\in P_{n},Or_{\rho}(\mu)\subset A, \label{inv2_}%
\end{equation}%
\begin{equation}
\forall\rho\in P_{n},\forall t\in\mathbf{R},\Phi^{\rho}(\mu,t)\in A,
\label{inv2__}%
\end{equation}%
\begin{equation}
\forall\alpha\in\Pi_{n},\forall k\in\mathbf{N},\Phi^{\alpha^{0}...\alpha^{k}%
}(\mu)\in A,
\end{equation}%
\begin{equation}
\forall\lambda\in\mathbf{B}^{n},\Phi^{\lambda}(\mu)\in A. \label{inv_1_}%
\end{equation}

\end{theorem}

\begin{proof}
(\ref{inv2__})$\Longrightarrow$(\ref{inv_1_}) Let $\mu\in A,\lambda
\in\mathbf{B}^{n}$ and the function $\rho\in P_{n}$ be arbitrary,%
\begin{equation}
\rho(t)=\alpha^{0}\cdot\chi_{\{t_{0}\}}(t)\oplus...\oplus\alpha^{k}\cdot
\chi_{\{t_{k}\}}(t)\oplus... \label{inv3}%
\end{equation}
with $\alpha\in\Pi_{n}$ and $(t_{k})\in Seq.$ We define%
\[
\rho^{\prime}(t)=\lambda\cdot\chi_{\{t^{\prime}\}}(t)\oplus\alpha^{0}\cdot
\chi_{\{t^{\prime}+t_{0}\}}(t)\oplus...\oplus\alpha^{k}\cdot\chi_{\{t^{\prime
}+t_{k}\}}(t)\oplus...
\]
where $t^{\prime}\in\mathbf{R}$ is arbitrary and we can see that $\rho
^{\prime}\in P_{n}$. (\ref{inv2__}) implies $\Phi^{\lambda}(\mu)=\Phi
^{\rho^{\prime}}(\mu,t^{\prime})\in A.$

(\ref{inv_1_})$\Longrightarrow$(\ref{inv2__}) Let $\mu\in A$ and $\rho\in
P_{n}$ be arbitrary, given by (\ref{inv3}), with $\alpha\in\Pi_{n},(t_{k})\in
Seq$. We get by induction on $k:$

$t<t_{0}:\qquad\qquad\Phi^{\rho}(\mu,t)=\mu\in A,$

$t\in\lbrack t_{0},t_{1}):\qquad\Phi^{\rho}(\mu,t)=\Phi^{\alpha^{0}}(\mu)\in
A$ from (\ref{inv_1_}),%
\[
...
\]

$t\in\lbrack t_{k-1},t_{k}):$ \ \ \ $\Phi^{\alpha^{0}...\alpha^{k-1}}(\mu)\in
A$ due to the hypothesis of the induction,

$t\in\lbrack t_{k},t_{k+1}):$ \ \ \ $\Phi^{\rho}(\mu,t)=\Phi^{\alpha^{k}}%
(\Phi^{\alpha^{0}...\alpha^{k-1}}(\mu))\in A$ from (\ref{inv_1_}),%
\[
...
\]
The rest of the implications are obvious.
\end{proof}

\begin{definition}
\label{Def166}The set $A\in P^{\ast}(\mathbf{B}^{n})$ is called a
\textbf{p-invariant} (or \textbf{p-stable}) \textbf{set} of the system
$\Xi_{\Phi}$ if it fulfills for any $\mu\in A$ one of (\ref{inv2}),...,
(\ref{inv3__}) and it is called an \textbf{n-invariant }(or n-\textbf{stable})
\textbf{set} of $\Xi_{\Phi}$ if it fulfills $\forall\mu\in A$ one of
(\ref{inv2_}),..., (\ref{inv_1_}).
\end{definition}

\begin{remark}
In the previous terminology, the letter 'p' comes from 'possibly' and the
letter 'n' comes from 'necessarily'. Both 'p' and 'n' refer to the
quantification of $\rho$. Such kind of p-definitions and n-definitions
recalling logic are caused by the fact that we translate 'real' concepts into
'binary' concepts and the former have no $\rho$ parameters, thus after
translation $\rho$ may appear quantified in two ways. The obvious implication
is n-invariance $\Longrightarrow$ p-invariance.
\end{remark}

\begin{example}
Let $\Phi:\mathbf{B}^{2}\rightarrow\mathbf{B}^{2}$ be defined by $\forall
\mu\in\mathbf{B}^{2},\Phi(\mu_{1},\mu_{2})=(\overline{\mu_{1}},\overline
{\mu_{2}})$ and $\rho(t)=(1,1)\cdot\chi_{\{0,1,2,...\}}(t).$ The set
$A=\{(0,1),(1,0)\}$ fulfills $\forall\mu\in A,\forall t\in\mathbf{R}%
,\Phi^{\rho}(\mu,t)\in A$ i.e. it satisfies ($\ref{inv1__}$):%
\[
\Phi^{\rho}((0,1),t)=(0,1)\cdot\chi_{(-\infty,0)}(t)\oplus(1,0)\cdot
\chi_{\lbrack0,1)}(t)\oplus
\]%
\[
\oplus(0,1)\cdot\chi_{\lbrack1,2)}(t)\oplus(1,0)\cdot\chi_{\lbrack
2,3)}(t)\oplus...
\]%
\[
\Phi^{\rho}((1,0),t)=(1,0)\cdot\chi_{(-\infty,0)}(t)\oplus(0,1)\cdot
\chi_{\lbrack0,1)}(t)\oplus
\]%
\[
\oplus(1,0)\cdot\chi_{\lbrack1,2)}(t)\oplus(0,1)\cdot\chi_{\lbrack
2,3)}(t)\oplus...
\]%
\begin{figure}
[ptb]
\begin{center}
\fbox{\includegraphics[
natheight=0.896000in,
natwidth=1.312800in,
height=0.9288in,
width=1.3474in
]%
{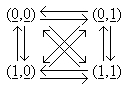}%
}\caption{The sets $\{(0,1),(1,0)\}$ and $\{(0,0),(1,1)\}$ are p-invariant.}%
\label{inv8}%
\end{center}
\end{figure}
see Figure \ref{inv8}; $A=\{(0,0),(1,1)\}$ satisfies the same invariance property.
\end{example}

\begin{example}
\label{Exa19}We define the function $\Phi:\mathbf{B}^{2}\rightarrow
\mathbf{B}^{2}$ by $\forall\mu\in\mathbf{B}^{2},$ $\Phi(\mu_{1},\mu_{2})$
$=(\mu_{1},\overline{\mu_{2}}),$ see Figure \ref{inv9}.
\begin{figure}
[ptb]
\begin{center}
\fbox{\includegraphics[
natheight=0.885600in,
natwidth=1.177000in,
height=0.9184in,
width=1.2107in
]%
{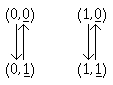}%
}\caption{The sets $\{(0,0),(0,1)\}$ and $\{(1,0),(1,1)\}$ are n-invariant.}%
\label{inv9}%
\end{center}
\end{figure}
We notice that the sets $A=\{(0,0),(0,1)\}$ and $A=\{(1,0),(1,1)\}$ are
n-invariant, as they fulfill $\forall\mu\in A,\forall\rho\in P_{2},Or_{\rho
}(\mu)=A.$
\end{example}

\begin{theorem}
\label{The81}Let be $\mu\in\mathbf{B}^{n}$ and $\rho^{\prime}\in P_{n}.$

a) If $\Phi(\mu)=\mu,$ then $\{\mu\}$ is an n-invariant set and the set $Eq$
of the fixed points of $\Phi$ is also n-invariant;

b) the set $Or_{\rho^{\prime}}(\mu)$ is p-invariant and $\underset{\rho\in
P_{n}}{\bigcup}Or_{\rho}(\mu)$ is n-invariant;

c) the set $\omega_{\rho^{\prime}}(\mu)$ is p-invariant.
\end{theorem}

\begin{proof}
a) From Corollary \ref{Cor17} we have that%
\[
\forall\rho\in P_{n},\forall t\in\mathbf{R},\Phi^{\rho}(\mu,t)=\mu\in\{\mu\}.
\]
Furthermore, we infer $\forall\mu^{\prime}\in Eq,$ $\forall\rho\in P_{n},$
$\forall t\in\mathbf{R},$%
\[
\Phi^{\rho}(\mu^{\prime},t)=\mu^{\prime}\in Eq.
\]

b) Let be $\mu^{\prime}\in Or_{\rho^{\prime}}(\mu),$ thus $t^{\prime}%
\in\mathbf{R}$ exists such that $\mu^{\prime}=\Phi^{\rho^{\prime}}%
(\mu,t^{\prime}).$ Then $\forall t\in\mathbf{R},$%
\[
\Phi^{\rho^{\prime}\cdot\chi_{(t^{\prime},\infty)}}(\mu^{\prime},t)=\left\{
\begin{array}
[c]{c}%
\Phi^{\rho^{\prime}}(\mu,t),t>t^{\prime}\\
\mu^{\prime},t\leq t^{\prime}%
\end{array}
\right.  \in Or_{\rho^{\prime}}(\mu).
\]
We have proved that $Or_{\rho^{\prime}}(\mu)$ is p-invariant.

We remark the equality%
\[
\underset{\rho\in P_{n}}{\bigcup}Or_{\rho}(\mu)=\underset{\alpha\in\Pi_{n}%
}{\bigcup}\{\Phi^{\alpha^{0}...\alpha^{k}}(\mu)|k\in\mathbf{N}\}
\]
and let us take an arbitrary $\mu^{\prime}\in\underset{\rho\in P_{n}}{\bigcup
}Or_{\rho}(\mu).$ If $\mu^{\prime}=\mu$ then the statement of the theorem is
proved, thus we can assume that $\mu^{\prime}\neq\mu,\mu^{\prime}=\Phi
^{\alpha^{0}...\alpha^{k}}(\mu),$ $\alpha^{0},...,\alpha^{k}\in\mathbf{B}%
^{n}.$ For any $\rho^{\prime\prime}\in P_{n},$%
\[
\rho^{\prime\prime}=\beta^{0}\cdot\chi_{\{t_{0}^{\prime}\}}\oplus
...\oplus\beta^{k}\cdot\chi_{\{t_{k}^{\prime}\}}\oplus...
\]
$\beta\in\Pi_{n},(t_{k}^{\prime})\in Seq$ and any $t\in\mathbf{R},$ we have
that $\Phi^{\rho^{\prime\prime}}(\mu^{\prime},t)$ is an element of the
sequence $\Phi^{\alpha^{0}...\alpha^{k}}(\mu),$ $\Phi^{\alpha^{0}...\alpha
^{k}\beta^{0}}(\mu),$ $...,$ $\Phi^{\alpha^{0}...\alpha^{k}\beta^{0}%
...\beta^{k^{\prime}}}(\mu),...$ where $\alpha^{0},...,\alpha^{k},\beta
^{0},...,\beta^{k^{\prime}},$ $...\in\Pi_{n}.$ The conclusion is that
$\Phi^{\rho^{\prime\prime}}(\mu^{\prime},t)\in\underset{\rho\in P_{n}}%
{\bigcup}Or_{\rho}(\mu).$

c) This is a consequence of Theorem \ref{The20_} e).
\end{proof}

\section{The basin of p-attraction and the basin of n-attraction}

\begin{theorem}
\label{The138}We consider the set $A\in P^{\ast}(\mathbf{B}^{n}).$ For any
$\mu\in\mathbf{B}^{n},$ the following statements are equivalent%
\begin{equation}
\exists\rho\in P_{n},\omega_{\rho}(\mu)\subset A, \label{sta9}%
\end{equation}%
\begin{equation}
\exists\rho\in P_{n},\exists t^{\prime}\in R,\forall t\geq t^{\prime}%
,\Phi^{\rho}(\mu,t)\in A, \label{sta3_}%
\end{equation}%
\begin{equation}
\exists\alpha\in\Pi_{n},\exists k^{\prime}\in\mathbf{N},\forall k\geq
k^{\prime},\Phi^{\alpha^{0}...\alpha^{k}}(\mu)\in A
\end{equation}
and the following statements are equivalent too%
\begin{equation}
\forall\rho\in P_{n},\omega_{\rho}(\mu)\subset A, \label{sta14}%
\end{equation}%
\begin{equation}
\forall\rho\in P_{n},\exists t^{\prime}\in R,\forall t\geq t^{\prime}%
,\Phi^{\rho}(\mu,t)\in A, \label{sta13}%
\end{equation}%
\begin{equation}
\forall\alpha\in\Pi_{n},\exists k^{\prime}\in\mathbf{N},\forall k\geq
k^{\prime},\Phi^{\alpha^{0}...\alpha^{k}}(\mu)\in A.
\end{equation}

\end{theorem}

\begin{proof}
(\ref{sta9})$\Longrightarrow$(\ref{sta3_}) We presume that (\ref{sta9}) is
true. Some $t^{\prime}$ exists with%
\[
\omega_{\rho}(\mu)=\{\Phi^{\rho}(\mu,t)|t\geq t^{\prime}\}
\]
and we conclude that $\forall t\geq t^{\prime},$%
\[
\Phi^{\rho}(\mu,t)\in\omega_{\rho}(\mu)\subset A.
\]
(\ref{sta3_})$\Longrightarrow$(\ref{sta9}) As $t^{\prime\prime}\in\mathbf{R}$
exists with%
\[
\omega_{\rho}(\mu)=\{\Phi^{\rho}(\mu,t)|t\geq t^{\prime\prime}\},
\]
from the truth of (\ref{sta3_}) we have that
\[
\omega_{\rho}(\mu)\subset\{\Phi^{\rho}(\mu,t)|t\geq\max\{t^{\prime}%
,t^{\prime\prime}\}\}\subset A.
\]

\end{proof}

\begin{definition}
\label{Def123}The \textbf{basin }(or \textbf{kingdom}, or \textbf{domain})
\textbf{of p-attraction} or the \textbf{p-stable set} of the set $A\in
P^{\ast}(\mathbf{B}^{n})$ is given by $\overline{W}(A)=\{\mu|\mu\in
\mathbf{B}^{n},\exists\rho\in P_{n},\omega_{\rho}(\mu)\subset A\};$ the
\textbf{basin }(or \textbf{kingdom}, or \textbf{domain}) \textbf{of
n-attraction} or the \textbf{n-stable set} of the set $A$ is given by
$\underline{W}(A)=\{\mu|\mu\in\mathbf{B}^{n},\forall\rho\in P_{n},\omega
_{\rho}(\mu)\subset A\}.$
\end{definition}

\begin{remark}
Definition \ref{Def123} makes use of the properties (\ref{sta9}) and
(\ref{sta14}). We can make use also in this Definition of the other equivalent
properties from Theorem \ref{The138}.

In Definition \ref{Def123}, one or both basins of attraction $\overline
{W}(A),\underline{W}(A)$ may be empty.
\end{remark}

\begin{theorem}
\label{The25}We have:

i) $\overline{W}(\mathbf{B}^{n})=\underline{W}(\mathbf{B}^{n})=\mathbf{B}%
^{n};$

ii) if $A\subset A^{\prime},$ then $\overline{W}(A)\subset\overline
{W}(A^{\prime})$ and $\underline{W}(A)\subset\underline{W}(A^{\prime})$ hold.
\end{theorem}

\begin{definition}
\label{Def168_}When $\overline{W}(A)\neq\emptyset,$ $A$ is said to be
\textbf{p-attractive} and for any non-empty set $B\subset\overline{W}(A),$ we
say that $A$ is \textbf{p-attractive} for $B$ and that $B$ is
\textbf{p-attracted} by $A$; $A$ is by definition \textbf{partially
p-attractive} if $\overline{W}(A)\notin\{\emptyset,\mathbf{B}^{n}\}$ and
\textbf{totally p-attractive} whenever $\overline{W}(A)=\mathbf{B}^{n}.$

The fact that $\underline{W}(A)\neq\emptyset$ makes us say that $A$ is
\textbf{n-attractive} and in this situation for any non-empty $B\subset
\underline{W}(A),$ $A$ is called \textbf{n-attractive} for $B$ and $B$ is
called to be \textbf{n-attracted} by $A;$ we use to say that $A$ is
\textbf{partially n-attractive} if $\underline{W}(A)\notin\{\emptyset
,\mathbf{B}^{n}\}$ and \textbf{totally n-attractive} if $\underline
{W}(A)=\mathbf{B}^{n}.$
\end{definition}

\begin{example}
We consider the system from Figure \ref{basin2}.
\begin{figure}
[ptb]
\begin{center}
\fbox{\includegraphics[
natheight=0.844100in,
natwidth=3.365000in,
height=0.8761in,
width=3.4108in
]%
{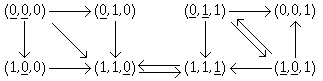}%
}\caption{Invariant sets and basins of attraction.}%
\label{basin2}%
\end{center}
\end{figure}
The set $A=\{(0,0,0)\}$ is neither p-invariant, nor n-invariant: $\overline
{W}(A)=\underline{W}(A)=\emptyset.$

The set $A=\{(0,0,0),(1,1,0),(1,1,1)\}$ is p-invariant but not n-invariant:
$\overline{W}(A)=\mathbf{B}^{3}\setminus\{(0,0,1)\},$ $\underline
{W}(A)=\{(0,0,0),(0,1,0),(1,0,0),$ $(1,1,0),(1,1,1)\}.$

We take $A=\{(1,1,0),(1,1,1),$ $(0,0,1)\}$ which is both p-invariant and
n-invariant. $A$ is totally p-attractive, $\overline{W}(A)=\mathbf{B}^{3}$ and
it is not totally n-attractive, since $\underline{W}(A)=\mathbf{B}%
^{3}\setminus\{(0,1,1),(1,0,1)\}.$

The set $A=\{(1,1,0),$ $(1,1,1),(0,1,1),(0,0,1),(1,0,1)\}$ is p-invariant,
n-invariant, totally p-attractive and totally n-attractive because
$\overline{W}(A)=\underline{W}(A)=\mathbf{B}^{3}.$
\end{example}

\begin{example}
The set $\mathbf{B}^{n}$ is totally p-attractive and totally n-attractive
(Theorem \ref{The25} i))$.$
\end{example}

\begin{theorem}
\label{The106}Let $A\in P^{\ast}(\mathbf{B}^{n})$ be some set. If $A$ is
p-invariant, then $A\subset\overline{W}(A)$ and $A$ is also p-attractive; if
$A$ is n-invariant, then $A\subset\underline{W}(A)$ and $A$ is also n-attractive.
\end{theorem}

\begin{proof}
Let $\mu\in A$ be arbitrary. The existence of $\rho\in P_{n}$ such that
$Or_{\rho}(\mu)\subset A$ (from the p-invariance of $A$) and the inclusion
$\omega_{\rho}(\mu)\subset Or_{\rho}(\mu)$ show that $\omega_{\rho}%
(\mu)\subset A,$ thus $\mu\in\overline{W}(A).$ As $\mu$ was arbitrary, we get
that $A\subset\overline{W}(A)$ and finally that $\overline{W}(A)\neq
\emptyset.$ $A$ is p-attractive.
\end{proof}

\begin{remark}
The previous Theorem shows the connection that exists between invariance and
attractiveness. If $A$ is p-attractive, then $\overline{W}(A)$ is the greatest
set that is p-attracted by $A$ and the point is that this really happens when
$A$ is p-invariant. The other situation is dual.
\end{remark}

\begin{theorem}
\label{The140}Let be $A\in P^{\ast}(\mathbf{B}^{n}).$ If $A$ is p-attractive,
then $\overline{W}(A)$ is p-invariant and if $A$ is n-attractive, then
$\underline{W}(A)$ is n-invariant.
\end{theorem}

\begin{proof}
If $A$ is p-attractive then $\overline{W}(A)\neq\emptyset$ and we prove that
$\overline{W}(A)$ is p-invariant. Let $\mu\in\overline{W}(A)$ be arbitrary and
fixed. From the definition of $\overline{W}(A)$ some $\rho\in P_{n}$ exists
with the property that $\omega_{\rho}(\mu)\subset A.$ We show that%
\[
\forall t^{\prime}\in\mathbf{R},\Phi^{\rho}(\mu,t^{\prime})\in\overline
{W}(A),
\]
i.e.%
\[
\forall t^{\prime}\in\mathbf{R},\exists\rho^{\prime}\in P_{n},\omega
_{\rho^{\prime}}(\Phi^{\rho}(\mu,t^{\prime}))\subset A.
\]
Indeed, we fix arbitrarily some $t^{\prime}\in\mathbf{R}$. With%
\[
\rho^{\prime}=\rho\chi_{(t^{\prime},\infty)}%
\]
we can write, from Remark \ref{Rem23}, equation (\ref{asterisc}) that%
\[
\omega_{\rho^{\prime}}(\Phi^{\rho}(\mu,t^{\prime}))=\omega_{\rho
\chi_{(t^{\prime},\infty)}}(\Phi^{\rho}(\mu,t^{\prime}))=\omega_{\rho}%
(\mu)\subset A.
\]

We prove now that $\underline{W}(A),$ which is non-empty from the
n-attractiveness of $A$, is also n-invariant. The property%
\[
\forall\mu^{\prime}\in\underline{W}(A),\forall\rho^{\prime}\in P_{n}%
,Or_{\rho^{\prime}}(\mu^{\prime})\subset\underline{W}(A),
\]
that is equivalent with%
\[
\forall\mu^{\prime}\in\underline{W}(A),\forall\rho^{\prime}\in P_{n}%
,\forall\mu^{\prime\prime}\in Or_{\rho^{\prime}}(\mu^{\prime}),\mu
^{\prime\prime}\in\underline{W}(A)
\]
and with%
\[
\forall\mu^{\prime}\in\mathbf{B}^{n},\forall\rho\in P_{n},\omega_{\rho}%
(\mu^{\prime})\subset A\Longrightarrow
\]%
\[
\Longrightarrow\forall\rho^{\prime}\in P_{n},\forall\mu^{\prime\prime}\in
Or_{\rho^{\prime}}(\mu^{\prime}),\forall\rho^{\prime\prime}\in P_{n}%
,\omega_{\rho^{\prime\prime}}(\mu^{\prime\prime})\subset A,
\]
means the following. Let $\mu^{\prime}\in\mathbf{B}^{n}$ and $\rho
^{\prime\prime}\in P_{n}$ be arbitrary and fixed. The hypothesis states that
for any%
\[
\rho=\alpha^{0}\cdot\chi_{\{t_{0}\}}\oplus...\oplus\alpha^{k}\cdot
\chi_{\{t_{k}\}}\oplus...
\]
$\alpha\in\Pi_{n},(t_{k})\in Seq$ we have%
\begin{equation}
\exists k_{1}\in\mathbf{N},\{\Phi^{\alpha^{0}...\alpha^{k}}(\mu^{\prime
})|k\geq k_{1}\}(=\omega_{\rho}(\mu^{\prime}))\subset A. \label{unu}%
\end{equation}
We consider arbitrarily the function $\rho^{\prime}\in P_{n},$%
\[
\rho^{\prime}=\alpha^{\prime0}\cdot\chi_{\{t_{0}^{\prime}\}}\oplus
...\oplus\alpha^{\prime k}\cdot\chi_{\{t_{k}^{\prime}\}}\oplus...
\]
$\alpha^{\prime}\in\Pi_{n},(t_{k}^{\prime})\in Seq$ and the point $\mu
^{\prime\prime}\in Or_{\rho^{\prime}}(\mu^{\prime}),$ thus $k^{\prime}%
\in\mathbf{N}$ exists with the property%
\[
\mu^{\prime\prime}=\Phi^{\alpha^{\prime0}...\alpha^{\prime k^{\prime}}}%
(\mu^{\prime}).
\]
We put $\rho^{\prime\prime}$ under the form%
\[
\rho^{\prime\prime}=\alpha^{\prime\prime0}\cdot\chi_{\{t_{0}^{\prime\prime}%
\}}\oplus...\oplus\alpha^{\prime\prime k}\cdot\chi_{\{t_{k}^{\prime\prime}%
\}}\oplus...
\]
$\alpha^{\prime\prime}\in\Pi_{n},(t_{k}^{\prime\prime})\in Seq$. The sequence%
\[
\Phi^{\alpha^{\prime\prime0}...\alpha^{\prime\prime k}}(\mu^{\prime\prime
})=\Phi^{\alpha^{\prime\prime0}...\alpha^{\prime\prime k}}(\Phi^{\alpha
^{\prime0}...\alpha^{\prime k^{\prime}}}(\mu^{\prime}))=\Phi^{\alpha^{\prime
0}...\alpha^{\prime k^{\prime}}\alpha^{\prime\prime0}...\alpha^{\prime\prime
k}}(\mu^{\prime}),
\]
$k\in\mathbf{N}$ fulfills the property (\ref{unu}), thus%
\[
\exists k_{2}\in\mathbf{N},\{\Phi^{\alpha^{\prime\prime0}...\alpha
^{\prime\prime k}}(\mu^{\prime\prime})|k\geq k_{2}\}(=\omega_{\rho
^{\prime\prime}}(\mu^{\prime\prime}))\subset A.
\]

\end{proof}

\begin{corollary}
If the set $A\in P^{\ast}(\mathbf{B}^{n})$ is p-invariant, then $\overline
{W}(A)$ is p-invariant and if $A$ is n-invariant, then the basin of
n-attraction $\underline{W}(A)$ is n-invariant.
\end{corollary}

\begin{proof}
These result from Theorem \ref{The106} and Theorem \ref{The140}.
\end{proof}

\section{Discussion}

Some notes on the terminology:

- universality means the greatest in the sense of inclusion. Any $X\subset
\Xi_{\Phi}$ is a system, but we did not study such systems in the present paper;

- regularity means the existence of a generator function $\Phi$, i.e.
analogies with the dynamical systems theory;

- autonomy means here that no input exists. We mention the fact that autonomy
has another non-equivalent definition also, a system is called autonomous if
its input set has exactly one element;

- asynchronicity refers (vaguely) to the fact that we work with real time and
binary values. Its antonym synchronicity means that 'discrete time' (and
binary values) in which the iterates of $\Phi$ are: $\Phi,\Phi\circ
\Phi,...,\Phi\circ...\circ\Phi,...$ i.e. in the sequence $\Phi^{\alpha^{0}%
},\Phi^{\alpha^{0}\alpha^{1}},...,\Phi^{\alpha^{0}...\alpha^{k}},...$ all
$\alpha^{k}$ are $(1,...,1),k\in\mathbf{N}.$ That is the discrete time of the
dynamical systems.

Our concept of invariance from Definition \ref{Def166} reproduces the point of
view expressed in \cite{bib1}, page 11, where the dynamical system
$S=(T,X,\Phi)$ is given, with $T=\mathbf{R}$ the time set, $X$ the state space
and $\Phi:T\times X\rightarrow X$ the flow: the set $A\subset X$ is said to be
invariant for the system $S$ if $\forall x\in A,\forall t\in T,\Phi_{t}(x)\in
A.$ This idea coincides with the one from \cite{bib2}, page 27 where the state
space $X$ is a differentiable manifold $M.$

In \cite{bib6}, page 92 the set $A\subset X$ is called globally invariant via
$\Phi$ if $\forall t\in T,\Phi_{t}(A)=A,$ recalling the situation of Example
\ref{Exa19} and Figure \ref{inv9}. In \cite{bib5}, page 3, the global
invariance and the invariance of $A\subset X$ are defined like at \cite{bib6}
and \cite{bib1}.

We mention also the definition of invariance from \cite{bib4}, page 19. Let
$P=(T,X,\Phi)$ be a process, where $T=\mathbf{R},$ $X$ is the state space and
$\Phi:\overline{T}\times X\rightarrow X$ is the flow of $P;$ we have denoted
$\overline{T}=\{(t^{\prime},t)|t^{\prime},t\in T,t\leq t^{\prime}\}.$ Then
$A\subset X$ is invariant relative to $\Phi$ if $\Phi_{t^{\prime},t}(A)\subset
A$ for any $(t^{\prime},t)\in\overline{T}.$ This last definition agrees itself
with ours in the special case when $t^{\prime}=0$ but it is more general since
it addresses systems which are not time invariant.

Stability is defined in \cite{bib2}, page 27 where $M$ is a differentiable
manifold and the evolution operator $\Phi_{t}:M\rightarrow M,t\in T$ is given.
The subset $A\subset M$ is stable for $\Phi$ if for any sufficiently small
neighborhood $U$ of $A$ a neighborhood $V$ of $A$ exists such that $\forall
x\in V,\forall t\geq0,\Phi_{t}(x)\in U.$ In our case when $M=\mathbf{B}^{n}$
has the discrete topology$,$ $A\subset\mathbf{B}^{n}$ and $U=V=A,$ this comes
to the invariance of $A$.

In \cite{bib1}, page 16 the closed invariant set $A\subset X$ is called stable
for $(T,X,\Phi)$ if i) for any sufficiently small neighborhood $U\supset A$
there exists a neighborhood $V\supset A$ such that $\forall t>0,\forall x\in
V,\Phi_{t}(x)\in U$ and ii) there exists a neighborhood $W\supset A$ such that
$\forall x\in W,\Phi_{t}(x)\rightarrow A$ as $t\rightarrow\infty.$ We see that
i) is the same request like at \cite{bib2} and ii) brings nothing new (item i)
means $Or_{\rho}(\mu)\subset A,$ thus a stronger request than item ii) which
is $\omega_{\rho}(\mu)\subset A$ in our case).

In a series of works (\cite{bib2}, page 27), either the set $A\subset M$ is
called asymptotically stable if it is stable and attractive, where $M$ is a
differentiable manifold, or (\cite{bib6}, page 112, \cite{bib5}, page 5) the
fixed point $x_{0}\in X$ is called asymptotically stable if it is stable and
attractive. We interpret stability as invariance and stating that $A$ or
$x_{0}$ is stable and attractive means that it is invariant and a weaker
property than invariance takes place (see Theorem \ref{The106}) and finally
asymptotic stability means invariance too.

In \cite{bib3}, page 132 the statement is made that many times, in
applications, by stability is understood attractiveness. This would mean, in
the conditions of Theorem \ref{The106}, weakening the invariance request and
we cannot accept this point of view.

In literature, \cite{bib3} defines at page 6 the basin of attraction of a
chaotic attractor $A\subset X$ as the set of the points whose $\omega-$limit
set is contained in $A$. This was reproduced at (\ref{sta9}) and
(\ref{sta14}), where $A\in P^{\ast}(\mathbf{B}^{n})$ was considered arbitrary however.

The work \cite{bib6} defines at page 124 the kingdom of attraction of an
attractive set $A\subset X$ as the greatest set of points of $X$ whose dynamic
ends (for $t\rightarrow\infty$) in $A$; when the kingdom of attraction is an
open set, it is called basin of attraction. For us, all the subsets
$A\subset\mathbf{B}^{n}$ are open in the discrete topology of $\mathbf{B}%
^{n}.$

In \cite{bib6}, page 123 the invariant set $A\subset X$ is called attractive
set for $B\subset X$ if the distance between $A$ and $\Phi_{t}(B)$ tends to 0
for $t\rightarrow\infty;$ a set $A$ is attractive if $B\neq\emptyset$ exists
that is attracted by $A$. A slightly different idea is expressed in
\cite{bib5}, page 4 where the invariant set $A$ is called attractive for $B$
if $\underset{t\rightarrow\infty}{\lim}\Phi_{t}(B)=A.$ Unlike these
definitions, in Definition \ref{Def168_} the set $A\subset\mathbf{B}^{n}$ is
not required to be invariant and the statement $B\subset\overline{W}(A)$
showing that $B$ is p-attracted by $A$, i.e. $\forall\mu\in B,\exists\rho\in
P_{n},\omega_{\rho}(\mu)\subset A,$ reproduces the fact that the distance
between $A$ and $\Phi_{t}(B)$ tends to 0 for $t\rightarrow\infty.$

In \cite{bib2}, page 27 $M$ is a differentiable manifold and the subset
$A\subset M$ is called attractive for $\Phi$ if a neighborhood $U$ of $A$
exists such that $\forall x\in U,\underset{t\rightarrow\infty}{\lim}\Phi
_{t}(x)\in A;$ in this case we say that $U$ is attracted by $A$. We have
reached (\ref{sta9}), (\ref{sta14}) and the requests of attractiveness
$\overline{W}(A)\neq\emptyset,\underline{W}(A)\neq\emptyset$ from Definition
\ref{Def168_}.

In \cite{bib3}, page 5 (Wiggins and Georgescu are cited) a closed invariant
set $A\subset X$ is called attractive if a neighborhood $U$ of $A$ exists such
that $\forall x\in U,\forall t\geq0,\Phi_{t}(x)\in U$ and $\Phi_{t}%
(x)\rightarrow A$ when $t\rightarrow\infty.$ Then the set $\underset{t\leq0}{%
{\displaystyle\bigcup}
}\Phi_{t}(U)$ is called the basin (the domain) of attraction of the set $A.$

In \cite{bib5}, page 4 the open set $W(A)\subset X$ representing the greatest
set of points of $X$ which is attracted by the attractive set $A$ is called
basin of attraction. This definition represents exactly $\overline
{W}(A),\underline{W}(A)$ from Definition \ref{Def123} in the circumstances
that (Definition \ref{Def168_}) the attractiveness of $A$ means that the
previous sets are non-empty.

We have the definition of the basin of attraction from \cite{bib2}, page 27:
the maximal set attracted by an attractor $A\subset X$ (invariant set,
attractive for one of its neighborhoods) is called the kingdom of attraction
of $A$; when the kingdom of attraction is an open set, it is called basin of
attraction. We conclude, related with the real to binary translation of this
definition, that if $A\in P^{\ast}(\mathbf{B}^{n})$ is p-invariant, then it is
p-attractive for itself and thus an 'attractor'; its basin of attraction
$\overline{W}(A)$ is non-empty in this case and it is the maximal set
attracted by $A$.

We note that the stable manifold of the equilibrium point $x_{0}\in X$ is
defined in \cite{bib5}, page 4 and \cite{bib6}, page 93 for the dynamical
system $(T,X,\Phi)$ by $W(x_{0})=\{x\in X|\underset{t\rightarrow\infty}{\lim
}\Phi_{t}(x)=x_{0}\}.$ In \cite{bib1}, page 46 the terminology of stable set
is used for this concept and \cite{bib5} mentions this terminology too. Thus,
by replacing $x_{0}\in X$ with $A\subset\mathbf{B}^{n}$ and $\underset
{t\rightarrow\infty}{\lim}\Phi_{t}(x)=x_{0}$ with $\omega_{\rho}(\mu)\subset
A$ we get for $\overline{W}(A),\underline{W}(A)$ the alternative terminology
of stable sets (i.e. invariant sets) of $A$.

\end{document}